\documentclass[12pt,preprint]{aastex}
\begin{document}

\title{Warm Molecular Hydrogen in the Galactic Wind of
  M82\footnote{This work is based on observations carried out with the
    facilities of NOAO.  NOAO is operated by the Association of
    Universities for Research in Astronomy (AURA), Inc. under
    cooperative agreement with the National Science Foundation.}}

\author{Sylvain Veilleux\altaffilmark{2,3}, David S. N. Rupke\altaffilmark{4}, 
and Rob Swaters\altaffilmark{2}}

\altaffiltext{2}{Department of Astronomy, University of Maryland,
  College Park, MD 20742; veilleux@astro.umd.edu,
  swaters@astro.umd.edu}

\altaffiltext{3}{Also: Max-Planck-Institut f\"ur extraterrestrische
  Physik, Postfach 1312, D-85741 Garching, Germany}

\altaffiltext{4}{Institute for Astronomy, University of Hawaii, 2680
  Woodlawn Drive, Honolulu, HI 96822; drupke@ifa.hawaii.edu}

\begin{abstract}
  We report the detection of a complex of extraplanar warm-H$_2$ knots
  and filaments extending more than $\sim$3 kpc above and below the
  galactic plane of M82, roughly coincident with the well-known
  galactic wind in this system. Comparisons of these data with
  published results at other wavelengths provide quantitative
  constraints on the topology, excitation, heating, and stability
  against disruption of the wind-entrained molecular ISM in this
  prototypical galactic wind.  Deep H$_2$ 2.12 $\mu$m observations
  such as these represent a promising new method to study the elusive
  but potentially important molecular component of galactic winds.
\end{abstract}


\clearpage

\section{Introduction}

Galaxy-scale outflows of gas (``superwinds'') are a ubiquitous
phenomenon in both starburst galaxies and those containing an active
galactic nucleus (Veilleux et al.\ 2005).
The observational data set on these outflows is steadily increasing,
but difficult issues remain. One vexing problem is how much different
phases of the ISM contribute to the mass and energy of
superwinds. Measurements have shown that winds contain cool (molecular
or neutral), warm (ionized) and hot (highly ionized) material.
However, the relative contribution of these phases to the total mass
and energy of the wind is uncertain by an order of magnitude. The
contribution from dust and molecular gas to the mass and energy in the
wind is almost completely unknown. The impact of superwinds on their
environments depends strongly on these quantities. Superwinds are
invoked as enrichers of galactic halos and the IGM, but it is not yet
clear if the winds extend far enough to carry dust, molecular gas, and
metals out of the galaxy.

Some evidence already exists for outflowing dust in galaxy halos.  A
few winds have been mapped in the far-infrared (Hughes et al.\ 1994;
Alton et al.\ 1999; Radovich et al.\ 2001; Kaneda et al.\ 2009), but
these maps are of low resolution and sensitivity and only trace the
coldest dust. In a few galaxies, large-scale, optically dark filaments
are observed (Phillips 1993; Cecil et al.\ 2001; Howk 2009). UV
observations have also revealed dust reflection in outflows in a few
galaxies (Hoopes et al.\ 2005). Neutral gas is outflowing in most
strong starbursts (Rupke et al.\ 2005a, 2005b) and extends over kpc
scales in a few resolved galaxies (Rupke et al.\ 2005b; Martin
2006). As shown by extinction measurements, dust correlates with the
neutral gas column in these galaxies, suggesting much of it is also
outflowing (Veilleux et al.\ 1995). Finally, spectacular dust
morphologies suggestive of large-scale winds have been detected in our
Galaxy based on data from the Midcourse Space Experiment (MSX)
satellite (Bland-Hawthorn \& Cohen 2003) and in the prototypical
outflow of M82 based on data obtained with {\em Spitzer} Infrared
Array Camera (IRAC; Engelbracht et al.\ 2006).  To our knowledge, M82
is the only object among all galaxies with dusty winds with an
unambiguously detected large-scale cold molecular outflow (Nakai et
al.\ 1987; Loiseau et al.\ 1990; Garc\'ia-Burillo et al.\ 2001; Walter
et al.\ 2002).

In this {\em Letter}, we present deep H$_2$ $v$ = 1$-$0 S(1) 2.12
$\mu$m images of M82 that reveal a complex of warm H$_2$-emitting
knots and filaments above and below the disk of M82, roughly
coincident with the ionized and dusty components of the galactic wind
in this object. The observations and methods of data reduction are
briefly described in Section 2. In Section 3, the results are
presented and compared with published data at other wavelengths. The
implications of these results on the topology, excitation, heating,
and stability of the wind-entrained material are discussed in Section
4. The main conclusions are summarized in Section 5. Throughout this
paper, we assume a distance of 3.53 Mpc for M82 (Karachentsev et al.\
2002).

\section{Observations \& Data Reduction}

The NOAO Extremely Wide Field Infrared Mosaic (NEWFIRM; Probst et al.\
2008 and references therein) on the Mayall 4-meter telescope at Kitt
Peak was used for the observations. This instrument images a
28$\arcmin$ $\times$ 28$\arcmin$ field of view with a 4K $\times$ 4K
pixel InSb array mosaic on a 0$\farcs$4 pixel scale. Total on-target
integrations of 40 and 420 minutes were obtained over a period of four
nights (1, 2, 4, and 5 November 2008 UT) using the
broadband K$_s$ and narrowband (1.1\% resolution) H$_2$ 2.124 $\mu$m
filters, respectively.  The so-called ``4Q'' observing mode of NEWFIRM
was used for efficient acquisition of the data: M82 was placed near
the center of one quadrant, and then it was cycled 10 (K$_s$) or 15
(H$_2$) times through each of the four quadrants, exposing at each
position for 10 or 60 seconds, respectively, and applying random
dither offsets within a box 30$\arcsec$ on a side so that the
positions did not exactly repeat from cycle to cycle.  The number of
frames coadded before being displayed was set to 6 at K$_s$ and 1
at H$_2$, to improve the acquisition efficiency of the K$_s$
observations.
Per-pixel digital averaging of 4 and high-order Fowler sampling
(Fowler \& Gatley 1990) of 8
were used to reduce read noise in the H$_2$ data.

NOAO NEWFIRM Science Pipeline v1.0 was used to reduce these data
(Swaters et al.\ 2009).  After the data were dark-subtracted,
linearized, and flat-fielded, a sky background image was determined
for each exposure by taking the median of the four preceding and four
subsequent exposures, but excluding exposures with the galaxy image in
it. This sky image was then scaled to match the sky background in the
galaxy exposures and subtracted. Astrometric solutions were obtained,
and the data were resampled and stacked. Next, the sky subtraction was
repeated, but while masking out objects detected in the first pass,
and a new stack was produced. Persistent images, bad pixels, and
transient effects were flagged and excluded in that stack.

\section{Results}

The continuum-subtracted H$_{2}$ image of M82 is shown in Figure
1. The stack of K$_s$ images was used, after proper intensity scaling,
to subtract off the continuum emission from the H$_2$ stack and
produce the ``pure'' H$_2$ image presented in this figure.
Immediately evident in Figure 1 are H$_{2}$ filaments extending more
than $\sim$3 kpc above and below the plane of the galaxy disk, roughly
coincident with the location of the galactic wind in M82 as traced by
the warm ionized gas (e.g.\ Shopbell \& Bland-Hawthorn 1998; Devine \&
Bally 1999; Lehnert et al.\ 1999; Westmoquette et al.\ 2009), the hot
ionized gas (e.g.\ Lehnert et al.\ 1999; Strickland \& Stevens 2000;
Stevens et al.\ 2003; Strickland \& Heckman 2009), the UV-scattering
dust (Hoopes et al.\ 2005), and the cold molecular gas (Walter et al.\
2002).

In Figures 2 and 3, our H$_2$ data are compared with the published
H$\alpha$ and 7.7 + 8.6 $\mu$m PAH images of Mutchler et al.\ (2007)
and Engelbracht et al.\ (2006), respectively. The H$\alpha$ data were
continuum-subtracted using an optimal linear combination of the V- and
I-band data. 
In the bright central disk (middle panels in Figure 3), there appears
to be large- and small-scale correspondence between the H$_2$ emission
and the PAH morphology.  Immediately outside this bright disk, the
correspondence is weaker: radial streamers are seen in both images but
their relative intensities differ with wavebands. None of these
streamers exactly coincides with the shock-excited [Fe~II] 1.64 $\mu$m
and mm-wave SiO filaments detected by Alonso-Herrero et al.\ (2003)
and Garc\'ia-Burillo et al.\ (2001), respectively.

On larger scales (top and bottom panels in Figure 3), the brightest
H$_2$ structures are near bright PAH structures. 
The ratio of H$_2$ to PAH emission in the bright H$_2$ clumps shows a
clear tendency to increase with distance from the nucleus (right panel
in Figure 2), while at fainter flux levels (left panel in Figure 2)
the PAH emission is overall more broadly distributed than the H$_2$
emission. This last result is not due to a lack of sensitivity at
H$_2$. It reflects a true physical difference in the distribution of
the warm H$_2$ gas relative to the faint PAH emission (see Section
4.2).

The H$\alpha$ emission tends to extend in roughly the same directions
as the H$_2$ and PAH emission, though the faint PAH and H$_2$
emissions are wider-angle.  The SE extension is prominent at all three
wavelengths, while the N-NE outflow is most prominent in
H$\alpha$. The H$_2$ emission in the SE extension is also fairly well
correlated with the ``inverted'' ionization cone seen in the
[N~II]/H$\alpha$ ratio map of Shopbell \& Bland-Hawthorn (1998; their
figure 4).  However, the match sometimes breaks down on smaller scale.
The examples shown in Figure 4 are not necessarily representative, but
they illustrate the complexity of the multi-phase material entrained
in the wind (see Sections 4.1 and 4.2).

One last but important comparison is made with the distributions of
the mm-wave CO emission from the cold molecular gas (Walter et al.\
2002) and the dust-scattered UV halo seen in the {\em GALEX} data
(Hoopes et al.\ 2005).  Most of the features seen in CO are detected
in H$_2$. The most prominent CO features are the central disk and the
``streamers'' that Walter et al.\ call S1-S4.  Two of these, S1 and
S2, are in the plane of the disk, and S1 at least is visible in H$_2$.
S3 and S4 extend to the E-SE; S3 corresponds to the more easterly
filament circled in the bottom panels of Figure 3. The CO emission in
the other H$_2$ filaments to the SE is very faint, if at all
present. As expected in photon-dominated regions (PDRs; Tielens \&
Hollenback 1985), the H$_2$ emission is more extended and probes
clouds which are more diffuse (smaller $A_V$) and at larger
distances from the nucleus than the CO emission. Deep H$_2$ 2.12
$\mu$m observations are therefore a promising new method to study the
molecular gas in galactic winds.

When combining all data sets, we find that the best multiwavelength
match is found in the S-SE outflow cone.  The UV emission does have
the filamentary emission features seen in H$_2$ and H$\alpha$.  In
fact, two of the three prominent H$_2$ filaments appear
well-correlated with the UV.  But, this is not true everywhere. The
E-SE S3 CO streamer, which is fairly prominent in H$_2$, is not seen
in UV or H$\alpha$ -- it appears to be obscured in the optical.  And,
in the NW region, CO is undetected, H$_2$ is faint, and the UV
emission almost seems anti-correlated to the H$_2$ emission. Complex
physical effects are clearly at work in this region. We return to this
issue in Section 4.2.

\section{Discussion}

\subsection{Entrainment of H$_2$}

The total amount of warm H$_2$ gas entrained in the wind of M82 may be
estimated from the luminosity of H$_2$ 2.12 $\mu$m outside the
disk. For this, we follow the calculations of Scoville et al.\ (1982)
which assume that the H$_2$ molecules are thermalized at T = 2000
K. The resulting prescription is $M_{H_2} =
0.00133~[L_{S(1)}/L_\odot]$ $M_\odot$. From our data, we find that
$M_{H_2} \sim 5000~M_\odot~{\rm (NW)} + 7000~M_\odot~{\rm (SE)} = 1.2
\times 10^4 M_\odot$ is located outside the central 5 $\times$ 1 kpc
region. This represents less than $\sim$ 10$^{-4}$ of the mass of
outflowing cold molecular material detected by Walter et al. (2002;
$\sim 4 \times 10^8 M_\odot$).  Assuming the warm H$_2$ material
shares the same kinematics as the cold molecular material, the total
kinetic energy of the warm H$_2$ material $\sim M_{H_2} v_{\rm
  H2-outflow}^2 \sim 10^{51} \times (v_{\rm H2-outflow}/v_{\rm
  CO-outflow})^2$ ergs, where $v_{\rm H2-outflow}$ is the average
H$_2$ outflow velocity and $v_{\rm CO-outflow} \sim 100$ km s$^{-1}$,
the average deprojected CO outflow velocity derived by Walter et al.\
(2002).  This is four orders of magnitudes lower than the kinetic
energies of the entrained ionized H$\alpha$-emitting gas (Shopbell \&
Bland-Hawthorn 1998, $\sim 2 \times 10^{55}$ ergs) and molecular
CO-emitting material (Walter et al.\ 2002, $\sim 3 \times 10^{55}$
ergs). The warm H$_2$ material is therefore not a dynamically
important component of the outflow.

The detailed processes by which the disk ISM is entrained in the wind
without destroying the molecular gas and mass-loading the wind in the
process are not well understood. The mere presence of molecular
material $\sim$3 kpc from the disk provides strong constraints on the
stability of wind-entrained clouds against photo- and
thermal-evaporation, Kelvin-Helmoltz instabilities, and shedding
events due to ablation (e.g.\ Marcolini et al.\ 2005).  The time
scale to bring such clouds out to a distance of 3 kpc is $\sim$ 10$^7$
$(v_{\rm H2-outflow}/v_{\rm CO-outflow})^{-1}$ yrs, assuming they
entered the wind near the center.  Recent high resolution
three-dimensional simulations of a non-uniform (fractal) radiative
cloud in a supersonic flow show that radiative cooling indeed helps
stabilize the cloud against disruption over a long enough time scale
to allow the cloud to reach velocities in excess of $\sim$ 100 km
s$^{-1}$, but it is not clear that the cloud can survive for as long
as $\sim$ 10$^7$ yrs in the flow (Cooper et al.\ 2009).

Numerical simulations tailored to the wind of M82 (Cooper et al.\
2008) show that the original distribution of the inhomogeneous ISM in
the disk is important in determining the overall morphology of the
wind, as well as the distribution of the entrained filaments. The
filled-in structures observed in the fractal ISM simulations of Cooper
et al.\ (2008) shares a greater resemblance with the complex
filamentary topology of the H$_2$ emission than the sharp-edge
quasi-conical structures derived from H$\alpha$ (McKeith et al.\ 1995;
Shopbell \& Bland-Hawthorn 1998; cf.\ Westmoquette et al.\ 2007,
2009). The dense molecular medium probed by H$_2$ 2.12 $\mu$m thus
keeps a stronger imprint of the multi-phase, cloudy ISM originally in
the disk than the ionized H$\alpha$- and X-ray-emitting medium, which
likely represents material that has broken up from the denser clouds
and has been accelerated further by the wind.

\subsection{Excitation and Heating of Extraplanar H$_2$}

There is a vast literature on the dense molecular gas in the disk of
M82. Molecular line studies indicate that the intense UV radiation
field from the starburst (large $G_0$ in the nomenclature of Tielens
\& Hollenbach 1985) has a strong influence on the physical conditions,
of the disk material (e.g.\ Lord et al.\ 1996; Mao et al.\ 2000;
Fuente et al.\ 2008; 
cf.\ Spaans \& Meijerink 2007 for a discussion of the effects of
X-rays). The near-infrared rovibrational H$_2$ emission lines are
produced in these PDRs (or XDRs, X-ray dominated regions, in Spaans \&
Meijerink 2007).

There are two basic ways to excite molecular hydrogen: collisional
excitation, i.e.\ inelastic collisions between molecules in a warm gas
($>$ 1000 K), or fluorescent excitation (``UV pumping'') through
absorption of soft-UV radiation (912 -- 1108 \AA) in the Lyman and
Werner bands. The latter dominates if $G_0/n$ is large ($n$ is the
hydrogen density in the PDR).  The method most commonly used to
differentiate between collisional excitation from fluorescence
consists in using flux ratios of various rovibrational H$_2$ lines
visible in the K band, particularly H$_2$ $v$ = 1$-$0 S(1) 2.12125
$\mu$m and the weaker $v$ = 1$-$0 S(0) 2.2227 $\mu$m and $v$ = 2$-$1
S(1) 2.2471 $\mu$m transitions. Results have favored thermal
excitation of H$_2$ and ruled out any significant radiative
fluorescent contribution in the {\em cores} of most starburst
galaxies, including M82 (e.g.\ Moorwood \& Oliva 1990; F\"orster
Schreiber et al.\ 2001 and references therein). The UV flux seen by
the halo material of M82 is necessarily lower than in the nucleus, but
the density of this material is probably also lower than in the disk,
so a significant contribution from UV excitation cannot be formally
ruled out in the extraplanar H$_2$ of M82.

In the alternative scenario of collisional excitation, three
mechanisms may provide the heating: (1) UV radiation from the
starburst, (2) X-rays from the starburst and wind, or (3) shocks
induced by the outflow. A strong constraint on the importance of UV
heating can be derived from the ratios of H$_2$ 2.12 $\mu$m to
H$\alpha$ (or, equivalently, Br$\gamma$; e.g.\ Puxley et al.\ 1990;
Doyon et al.\ 1994), while the relative H$_2$ 1$-$0 S(1) and X-ray
fluxes provide an excellent way to test the mechanism of X-ray heating
(e.g.\ Veilleux et al.\ 1997). The third and final scenario may be
tested using shock diagnostics. The lack of one-to-one match between
the radial streamers seen near the disk in H$_2$ (Figure 3, middle)
and the shock-heated [Fe~II] and SiO filaments of Alonso-Herrero et
al.\ (2003) and Garc\'ia-Burillo et al.\ (2001) favors UV
excitation/heating or X-ray heating over shock heating in this region.
The fact that the brightest H$_2$ emission in the SE extension is
fairly well correlated with the ``inverted'' ionization cone seen in
the [N~II]/H$\alpha$ ratio map of M82 suggests that UV
excitation/heating is important there, since this is a region where
photoionization by OB stars dominates over shocks (Shopbell \&
Bland-Hawthorn 1998). One also expects a loose correlation between
H$_2$ and PAH emission in this region since the latter is due to
transient heating by individual near-UV ($<$ 13.6 eV) photons; this is
confirmed in Figure 3 (bottom panels).

Outside this region, shocks are believed to be relatively more
important at producing the bright H$\alpha$ filaments (larger
[N~II]/H$\alpha$ ratios are observed there; Shopbell \& Bland-Hawthorn
1998). This may explain the larger H$_2$-to-PAH ratios there (enhanced
H$_2$ emission from shock heating) and generally weaker correspondence
between H$_2$, PAH, and H$\alpha$ features (H$\alpha$ emission
requires energetic $>$ 13.6 eV photons or shocks which are capable of
destroying H$_2$ and macromolecules like PAHs; e.g.\ Reach et al.\
2006). However, this explanation is clearly not valid in the fainter,
more diffuse, PAH-emitting material since it is not detected in
H$\alpha$ and barely visible in H$_2$.  The data do not allow us to
determine whether the conditions in this gas favor higher UV
photodissociation of H$_2$ (e.g.\ lower H$_2$ self-shielding due to
broader lines), higher near-UV heating of PAHs, or lower H$_2$
formation rates on grains (e.g.\ lower density gas, higher grain
processing, or lower sticking coefficient in warm gas; Wolfire et al.\
2008 and references therein).

\section{Concluding Remarks}

We have shown that deep H$_2$ 2.12 $\mu$m observations are an
excellent complement to mid-infrared PAH and mm-wave CO observations
in the search and analysis of the dusty molecular component in the
winds of nearby galaxies. Our deep H$_2$ 2.12 $\mu$m image of M82
reveals a complex of knots and filaments that extends more than
$\sim$3 kpc above and below the disk plane in the same general
direction as the well-known galactic wind in this system. This warm
molecular material is not a dynamically important component of the
outflow, but it is potentially a sensitive tracer of the cooler, more
dominant, molecular wind-entrained material.  Detailed morphological
comparisons with published data at other wavelengths reveal the
complex, multi-phase nature of the wind-entrained material. The
results favor UV excitation/heating (shock heating) as the principal
H$_2$ emission process in the inner (outer) bright filaments of the
wind, but other processes are probably at work in the fainter, more
diffuse, PAH-emitting material where the H$_2$ emission is apparently
suppressed relative to the PAHs. On-going and planned infrared
spectroscopy of M82 by various groups should soon be able to test
these results and better quantify the energetics, hence impact, of the
molecular wind on the large-scale environment of M82.

\acknowledgements Support for this work was provided to S.V. and
D.S.N.R. by NSF through contract AST 0606932.  S.V. also acknowledges
support from a Senior Award from the Alexander von Humboldt Foundation
and thanks the host institution, MPE Garching, where this paper was
written.  The authors are grateful to Ron Probst and the rest of the
NEWFIRM team for putting together such an outstanding instrument, and
to Frank Valdes and Tracy Huard for their assistance with the NEWFIRM
Science Pipeline.  The authors thank David Fanning for the expert
advice and software dealing with images in IDL, provided at {\tt
  www.dfanning.com}, and Mark Wolfire and the referee for a critical
reading of the manuscript and for making some key suggestions which
improved the paper.
This work has made use of NASA/ADS Abstract Service and the NASA/IPAC
Extragalactic Database (NED), which is operated by JPL/Caltech, under
contract with NASA.

\clearpage

\begin{figure*}
\epsscale{1.05}
\caption{H$_2$ 2.12 $\mu$m emission in M82. ({\em left}) ``Pure''
  H$_2$ emission on a false-color scale. ({\em right}) H$_2$ (red) +
  K$_s$ continuum (blue) emission.  The H$_2$ data were continuum
  subtracted and smoothed with a 4$\arcsec$ Gaussian kernel to bring
  out the large scale structure. The intensity scalings in this figure
  are ``asinh'' from Lupton et al.\ (1999). The flux scale is in ergs
  s$^{-1}$ cm$^{-2}$ arcsec$^{-2}$. }
\label{fig:m82-h2}
\end{figure*}

\begin{figure*}
\epsscale{1.05}
\caption{({\em left}) Three-color composite of M82: {\em Green:} H$_2$
  2.12 $\mu$ emission (same as Figure 1), {\em Red:} 7.7 + 8.6 $\mu$m
  PAH emission from Engelbracht et al.\ (2006), and {\em Blue:} {\em
    HST}/ACS continuum-subtracted H$\alpha$ emission from Mutchler et
  al.\ (2007), smoothed to 0$\farcs$2. See Figure 1 for information on
  intensity scaling. ({\em right}) H$_2$-to-PAH emisssion ratio map in
  the brightest H$_2$ filaments. Black regions are of low S/N or
  affected by saturation effects. The absolute ratio scale is accurate
  to within a factor of only $\sim$2 due to PSF mismatch between the
  two wavebands, lack of color corrections for the IRAC photometry,
  and stellar contamination to the 8 $\mu$m flux.  }
\label{fig:m82-h2-ha-pah1}
\end{figure*}

\begin{figure*}
\epsscale{1.0}
\caption{ Large-scale images (2$\arcmin$ wide) that focus on three
  regions of coherent H$_2$ 2.12 $\mu$m emission (central column;
  smoothed with 4$\arcsec$ Gaussian kernel) and compare with the
  H$\alpha$ (left; 0$\farcs$2) and PAH (right) emission. Shown in the
  top, middle, and bottom rows are the NE quadrant, nuclear disk
  region, and SE quadrant, respectively.  See Figure 1 for information
  on intensity scaling. The ellipses delineate prominent radial
  filaments that appear in at least one waveband.}
\label{fig:m82-h2-ha-pah2}
\end{figure*}

\begin{figure*}
\epsscale{0.9}
\caption{ Examples of small-scale features in H$_2$ 2.12 $\mu$m
  (middle row; 2$\arcsec$ Gaussian kernel), H$\alpha$ (top row;
  0$\farcs$2.) and PAH (bottom row) emission. See Figure 1 for
  information on intensity scaling. The circles are 10$\arcsec$ in
  diameter. }
\label{fig:m82-h2-ha-pah3}
\end{figure*}

\end{document}